\documentclass[aps,twocolumn,showpacs,amssymb,pra,10pt]{revtex4-1}
\usepackage{amsmath}
\usepackage{graphicx}

\usepackage[caption=false]{subfig}

\usepackage{bbm}
\usepackage{color}

\newcommand{\rmi}{\textrm{i}}
\newcommand{\rmd}{\textrm{d}}

\DeclareMathOperator{\im}{Im}
\DeclareMathOperator{\re}{Re}

\begin{document}
\title{Dynamical correlation functions for the one-dimensional Bose--Hubbard insulator}

\author{Kevin zu M\"unster}
\author{Florian Gebhard}
\affiliation{Fachbereich Physik, Philipps Universit\"at Marburg,
D-35032 Marburg, Germany}

\author{Satoshi Ejima}
\author{Holger Fehske}
\affiliation{Institut f\"ur Physik,
             Ernst-Moritz--Arndt Universit\"at Greifswald,
             D-17489 Greifswald,
             Germany}

\begin{abstract}%
We calculate the dynamical current and kinetic-energy correlation functions for the 
first Mott lobe of the one-dimensional Bose--Hubbard model.
We employ the strong-coupling expansion up to sixth order in $x=t/U$, and
the dynamical density-matrix renormalization group method on rings
with 64 sites. The correlation functions are finite above the single-particle gap
with a square-root onset, as is also found from field theory close to the Mott transition. 
The correlation functions display
a featureless superposition of the primary and tertiary Hubbard bands. 
We find very good agreement between all methods in the interaction/frequency regimes 
where they are applicable. 
\end{abstract}



\maketitle

\section{Introduction}
\label{sec:I}
Ultracold atomic systems in which the atoms are trapped
on an optical lattice offer a wide range of possibilities
to prove the applicability of theoretical studies~\cite{RevModPhys.80.885}. New
experimental techniques such as the modulation of the
amplitude and the phase of the lattice potential provide a
variety of new possibilities, e.g., the modulation
of the lattice potential can be used to introduce artificial
gauge fields into the system~\cite{RevModPhys.83.1523,PhysRevLett.108.225304}. Such a modulation
derives an excitation of the system described
by the kinetic and current correlation functions.
Recently, the signature of a Higgs amplitude mode in the
two-dimensional superfluid in the vicinity of a quantum
phase transition to a Mott insulator has been
predicted by field theory combined with quantum Monte Carlo
simulations, showing a resonance-like feature in
dynamical spectra [4]. Albeit this increasing attention,
theoretical studies for these dynamical correlation functions
in the bosonic systems are rare even in one-dimension.
K\"uhner {\it et al.} computed the optical conductivities
in the one-dimensional Bose--Hubbard model, using the density matrix
renormalization group (DMRG) method~\cite{KWM00}.
The dynamical response due to the lattice modulation
were also simulated applying the time-dependent DMRG~\cite{PhysRevA.84.043608}.

In this work we address the
one-dimensional Bose--Hubbard model in the insulating phase 
when the number of bosons~$N$ equals the number of lattice sites~$L$.
The model describes a system of neutral spin-less atoms trapped in an optical 
lattice with deep lattice potentials.
The Bose--Hubbard Hamiltonian is defined by
\begin{eqnarray}
\hat H&=& -t \hat{T} + U \hat{D} \nonumber \; ,\\
\hat{T} &=& \sum_{j=1}^L( \hat{b}_j^\dagger \hat{b}_{j+1}^{\phantom{\dagger}}
          +\hat{b}_{j+1}^{\dagger}\hat{b}_{j}^{\phantom{\dagger}} )
  \label{hamil} \; ,\\
\hat{D} &=& \frac{1}{2}\sum_{j=1}^{L} \hat{n}_j(\hat{n}_j-1)\; .
\nonumber
\end{eqnarray}
Here, $\hat{b}_j^{\dagger}$, $\hat{b}_j^{\phantom{\dagger}}$ and 
$\hat{n}_j=\hat{b}_j^\dagger \hat{b}_j^{\phantom{\dagger}}$ 
are the boson creation, annihilation, and particle number operators on 
site $j$, respectively, and periodic boundary conditions (PBC) apply.
Throughout this work, we denote the ratio of 
the hopping amplitude $t$ and the local interaction strength~$U$ by $x=t/U$.

At integer filling, $\rho=N/L=1$, the Bose--Hubbard model describes a phase 
transition from the Mott insulating phase to the superfluid phase at a critical 
interaction ratio, $x_{\rm c} \approx 0.305$~\cite{Ejima2011}. 
The phase transition lies in the XY universality class and is of Kosterlitz-Thouless type 
which results in an exponentially small band gap near the transition. 
In this work we restrict ourselves to the first Mott lobe, $x<x_{\rm c}$.

The response of the bosons to weak phase and amplitude modulations 
of the lattice potential is described
by the correlation functions for the current operator and the kinetic-energy operator;
for a detailed discussion of the connection between measurable 
quantities like the energy absorption rate and correlation functions 
in terms of a linear response theory, 
see Refs.~\cite{TG11,Iucci2006,Kollath2006}. 
We define the correlation function of an operator $\hat{A}$ as the imaginary 
part of the corresponding retarded Green function, which is given in frequency 
space by
\begin{eqnarray}
 \widetilde{S}_A(\omega\geq 0) &=& - \frac{1}{\pi}\lim_{\eta \to 0^+}  
\im \left[G^{\rm ret}_A(\omega+\rmi \eta) \right] 
\nonumber \\
 &=& \frac{1}{L} \sum_n |\langle \Psi_n |\hat{A} |\Psi_0 \rangle|^2 
\delta(\omega-(E_n-E_0)) \,. \label{eq:defSA}
\end{eqnarray}
Here, $|\Psi_0\rangle$ is the ground state of $\hat{H}$ with energy $E_0$,
and $|\Psi_n\rangle$ are eigenstates of $\hat{H}$ with energy $E_n$.
Note that the correlation functions are positive, $\widetilde{S}_A(\omega)\geq 0$.
Moreover, the sum rule
\begin{equation}
\int_0^{\infty} {\rm d} \omega \widetilde{S}_A(\omega) = \frac{1}{L}
\langle \Psi_0 |\hat{A}^2 |\Psi_0\rangle 
\label{eq:sumrule}
\end{equation}
shows that $\widetilde{S}_A(\omega\to\infty)\to 0$.

In this work, we focus on the correlation function for the kinetic energy, 
$\widetilde{S}_{T}(\omega)$ 
with $\hat{T}$ from Eq.~(\ref{hamil}) and define
\begin{equation}
\widetilde{S}_T(\omega) = w_0^T\delta(\omega) + S_T(\omega)
\end{equation}
where we extracted the $\delta$-peak at $\omega=0$.
Moreover, we address $S_{J}(\omega)$ with the current operator
\begin{eqnarray}
 \hat{J} &=& \rmi \sum_l\left( c_{j+1}^{\dagger} c_j^{\phantom{\dagger}} 
- c_{j}^{\dagger} c_{j+1}^{\phantom{\dagger}}\right) \;.
\end{eqnarray}
Note that $S_J$ is related to the real part 
of the optical conductivity at zero momentum as~\cite{mahan2000}
\begin{equation}
S_J(\omega>0) = \frac{\omega}{\pi} \re \left[ \sigma(q=0,\omega)\right]\;.
\end{equation}
Since we focus on the insulating phase, we know that $w_0^J=0$
so that $\widetilde{S}_J(\omega)=S_J(\omega)$.

Our work is organized as follows. In Sect.~\ref{sec:methods} we briefly discuss
the dynamical DMRG (DDMRG) method 
and the strong-coupling expansion, and give the field-theory 
expression for the current correlation function.
In Sect.~\ref{sec:cfs} we discuss 
the results for the current and kinetic-energy correlation functions in the Mott phase.
Conclusions and outlook, Sect.~\ref{sec:outlook}, close our presentation.
Some technical aspects are deferred to the Appendix.

\section{Methods}
\label{sec:methods}

We evaluate the correlation functions with the numerical DDMRG method
and the analytical strong-coupling expansion (SC). 
We used these methods previously~\cite{Ejima2012,EFG12,Ejima2013} 
for the evaluation of single-particle and two-particle
response functions such as the density 
correlation function $S_n(k,\omega)$. In Refs.~\cite{Ejima2012,Ejima2013}
a detailed explanation of the methods can be found. 
In this section, we briefly summarize the most important aspects of both methods.
For comparison, we also give the field-theoretical expression for the current
correlation function.
For an early application of the DMRG correction vector method~\cite{KW99}
to the one-dimensional (extended) Bose--Hubbard model, see Ref.~\cite{KWM00}.

\subsection{DDMRG}
\label{subsec:DDMRG}

In order to simulate the Bose--Hubbard type models using 
the (D)DMRG technique~\cite{Wh92,Je02b,JF07}, the maximum number of bosons per site $n_b$ 
should be limited, while each lattice site can be occupied in principle by 
infinitely many bosons. Nevertheless, the (D)DMRG results are unbiased
and numerically exact as long as the dependence on $n_b$ can be verified
to be negligible. 
In the DDMRG scheme it makes a considerable difference from the
correction vector method~\cite{KW99} to minimize the functional~\cite{Je02b}
\begin{eqnarray}
 W_{A,\eta}(\omega,\psi)
  &=&\langle\psi|(E_0+\omega-\hat{H})^2+\eta^2|\psi\rangle 
\nonumber \\
 &&+\eta\langle A|\psi\rangle +\eta\langle \psi|A \rangle,
 \label{eq:dmrg1}
\end{eqnarray} 
where $|A\rangle=\hat{A}|\psi_0\rangle$. Then the imaginary part of 
the Green function can be evaluated as
\begin{eqnarray}
 W_{A,\eta}(\omega,\psi_{\rm min})=-\pi\eta\im G_A^{\rm ret}(\omega+\rmi \eta).
 \label{eq:dmrg2}
\end{eqnarray}
Within this DDMRG scheme we repeat the sweeps in the finite-system
algorithm until the functional $W_{A,\eta}(\omega,\psi)$ takes
its minimal values.

Calculating the dynamical current [$A=J$ in Eq.~(\ref{eq:dmrg1}) and (\ref{eq:dmrg2})] and
kinetic-energy [$A=T$] correlation functions
in the first Mott lobe 
of the Bose--Hubbard model,
we keep $m=800$ states to determine the ground state during 
the first five DMRG sweeps, and then use $m=400$ states for 
the evaluation of the dynamical properties. 
For a precise analysis of the dynamical properties we consider
 a finite system with broadening width $\eta$ as small
as possible. In doing so artificial peaks can appear if $\eta$
is too small for fixed system size $L$.  Thus, one needs to find 
an appropriate $\eta(L)$ empirically. In order to avoid artificial peaks, in this 
paper we fix $\eta=0.8 t$ for $L=64$.

\subsection{Spectral broadening and deconvolution}
\label{subsec:broadening}

The DDMRG method works with complex frequencies, i.e., instead of
the real frequency $\omega$, the calculations are done
with a finite shift~$\eta$ into the complex plane,
$\omega \to \omega + \rmi \eta$. The shift to the complex plane
introduces a Lorentzian 
spectral broadening of the correlation function, i.e., the DDMRG actually provides
\begin{equation}
S_{A}^{\eta}(\omega) = \frac{1}{\pi}\int_{-\infty}^{\infty}
\frac{\eta}{(\omega-\widetilde{\omega})^2+\eta^2} S_{A}(\widetilde{\omega}) 
\rmd\widetilde{\omega}
\label{eq:convolutionintegral}
\end{equation}
at equally spaced frequencies $\omega_i$ with high numerical accuracy.
The size of the intervals 
$\Delta\omega=|\omega_{i+1}-\omega_i|$ is smaller than, but
of the order of $\eta$.
Then, without loss of accuracy, 
the integration over $\omega$ in~(\ref{eq:convolutionintegral}) 
can be represented by a matrix multiplication,
\begin{eqnarray}
S_{A}^{\eta}(\omega_i) 
&=& \sum_j 
\frac{1}{\pi}\frac{\eta}{(\omega_i-\omega_j)^2+\eta^2}
S_{A}(\omega_j) \;,
\nonumber \\
\underline{S_A^{\eta}}
 &=& \underline{\underline{L^{\eta}}}\cdot \underline{S_A} \; .
\label{eq:deconvolution-matrix}
\end{eqnarray}
The derivation of  the vector $\underline{S_A}$ for the correlation function
at the frequencies $\omega_i$ 
from the corresponding DDMRG vector $\underline{S_A^{\eta}}$ 
(`deconvolution') poses an ill-conditioned inverse problem.

There is a number of deconvolution techniques for Lorentz-broadened spectra.
It was shown in Refs.~\cite{GJMNN03,Nishimoto2004,PhysRevB.89.195101}
that the matrix inversion of~(\ref{eq:deconvolution-matrix})
provides a simple and reliable way to deconvolve spectral functions,
\begin{equation}
\underline{S_A} = 
\left(\underline{\underline{L^{\eta}}}\right)^{-1}\cdot \underline{S_A^{\eta}} \; .
\label{eq:decon}
\end{equation}
As seen from~(\ref{eq:defSA}), the correlation functions are positive,
$S_A(\omega)\geq 0$. However, the deconvolution scheme~(\ref {eq:decon})
cannot guarantee this so that 
the deconvolved correlation functions might be negative in some regions. 
The width and depth of the regions with negative values of the correlation
 functions can be taken as a sign of the error introduced by the deconvolution technique and 
the finite-size/finite-$\eta$ limitations in the DDMRG method.
Of course, the deviations are most prominent when the correlation
functions tend to zero or show narrow extrema. In the first Mott lobe
of the Bose-Hubbard model, the width of the peaks is large enough to deconvolve 
the data using $\eta=0.8t$ and $L=64$.   
Then, only very close to $\omega\simeq \Delta$ (with particle gap $\Delta$), the spectra show negative values
as we see in the following.

\subsection{Strong-coupling expansion}
\label{subsec:scexpansion}

In the strong-coupling expansion we use a Harris--Lange transformation~\cite{PhysRev.157.295,PhysRevB.49.7904} to 
obtain an 
effective Hamilton operator $\hat{h}$ that does not mix states from different 
subspaces of the operator for potential energy, $U \hat{D}$. 
Within this approach, the transformation operator $\hat{S}$ as well as 
the effective Hamilton operator $\hat{h}$ are expanded in $x$,
\begin{eqnarray}
\hat{h}&=&e^{\hat{S}} \hat{H} e^{-\hat{S}} 
= U \hat{D} +t \sum_{r=0}^{\infty}x^r \hat{h}_r \; , \nonumber \\ 
\hat{S}&=&-\hat{S}^{\dagger}=
\sum_{r=1}^{\infty}x^r \hat{S}_r \;. 
\label{eq:harris_lange_transformation}
\end{eqnarray}
In $m$th order SC perturbation theory,
the operators $\hat{h}_r$ and $\hat{S}_r$ are constructed 
iteratively whereby we enforce $[\hat{h}_r,\hat{D}]_-=0$ for $0\leq r\leq m$.
This requires that we keep the operators 
up $\hat{S}_{m}$, $\hat{h}_{m-1}$ in the Taylor series for $\hat{S}$ and $\hat{h}$.

The (non-degenerate) ground state $|\Psi_0\rangle$ for finite~$x$ 
is obtained from the (non-degenerate) ground state $|\Phi_0\rangle$ for $x=0$ as
\begin{equation}
|\Psi_0 \rangle = \exp (-\hat{S}) |\Phi_0 \rangle \; .
\label{eq:PhifromPsi}
\end{equation}
Note that the ground state for $t=0$ is very simple because
for $N=L$ every site is singly occupied by a boson,
$|\Phi_0 \rangle=\prod_j \hat{b}_j^+ |{\rm vac}\rangle$.
Within SC perturbation theory,
the evaluation of a ground state expectation 
values of an operator $\hat{A}$ 
reduces to
\begin{eqnarray}
\langle \Psi_0 | \hat{A} | \Psi_0\rangle &=& 
\langle \Phi_0 | \tilde{A} | \Phi_0\rangle \nonumber \; , \\
\tilde{A}&=&\exp(\hat{S})\hat{A}\exp(-\hat{S}) \; .
\end{eqnarray}
To lowest order, it is readily deduced from Eq.~(\ref{eq:PhifromPsi}) that
the operator $\hat{S}_1$ generates states in $|\Phi_0\rangle$
with one doubly occupied site and a neighboring hole.
In general, the states from the subspace $\mathcal{H}_1 \subset \mathcal{H}$ 
give rise to the primary Hubbard band
in the dynamical correlation functions around $\omega \approx U$.
In higher orders in~(\ref{eq:PhifromPsi}), 
states from $\mathcal{H}_2$ and $\mathcal{H}_3$ 
appear in $|\Psi_0\rangle$ that give rise to the 
secondary and tertiary Hubbard bands.
Therefore, as long as $x$ is small and 
the spectral splitting of the Hilbert space is large,
the dynamical correlation functions display contributions
from energetically separated Hubbard bands.
Due to sum rules and the fact that the correlation functions are nonnegative,
most of the spectral weight is concentrated in the primary Hubbard band.
Within SC perturbation theory, the contributions from the higher Hubbard bands 
is fairly small.

For the primary Hubbard band in $m$th order approximation, 
we have to consider one double occupancy and one hole
on a ring. When their
relative distance is larger than $m$ lattice sites,
they do not interact with each other. For this reason,
the SC problem reduces to the analytical solution of a two-particle problem
with a finite-range interaction~\cite{Ejima2012}. 
Most importantly, the interaction leads to a backscattering of the hole from the double 
occupancy so that an effective single-particle problem results 
where the hole moves on a chain with open boundary conditions.

For higher order corrections the analytical expressions for the weights and the Hamilton 
operator are employed for the succeeding numerical evaluation of the correlation functions. If the 
states describe a single quasiparticle, Cauchy's integral formula can be used to compute
the results for an infinite system \cite{Ejima2012,Ejima2013} without spectral broadening, $\eta=0^+$.
The contributions originating from states with two or more quasiparticles can be obtaine by a simple diagonalization 
of the Hamilton operator for systems large enough to set $\eta \approx 0$.
However, since this perturbation theory is based on the spectral separation 
of the Hilbert space, it breaks down when the gap becomes
exponentially small close to the Kosterlitz-Thouless type Mott transition
at $x_{\rm c}\approx 0.305$. Therefore, its applicability is limited
to $x\lesssim 0.10$, as we shall see below.

\subsection{Results from field theory}
\label{subsec:FT}

Close to the Mott transition, the Bose--Hubbard model
can be described by the sine-Gordon model where the
dispersion of holons (${\rm h}$) 
and antiholons ($\overline{\rm h}$) with momentum $P$
is given by $E(P)=\pm \sqrt{P^2+(\Delta/2)^2}$
where $\Delta$ is the (exponentially small) single-particle gap.
If only one holon and antiholon is taken into account and a marginal interaction between them is assumed,
the two-particle contribution to the current-current correlation function is 
given by~\cite{KS78,Smirnov92,CET01}
($\nu=\omega/\Delta$)
\begin{eqnarray}
S_{J,{\rm h},\overline{\rm h}}(\omega)
&=& \frac{\Delta}{\pi} C_2(\Delta)
\frac{2}{\pi}\frac{\sqrt{\nu^2-1}}{\nu}
\Theta(\nu-1)\label{2part}\\
&&
\exp\left(-\int_0^\infty\frac{{\rm d}x}{x}\frac{1-\cos(x\theta/\pi)\cosh x
}{
\exp(x/2)\cosh(x/2)\sinh x}\right) \; ,
\nonumber
\end{eqnarray}
where $\theta=2 {\rm Arccosh} (\nu)$. Formula~(\ref{2part})
is exact in the interval $\Delta\leq \omega\leq 2\Delta$. 
For $\omega >2\Delta$ there are corrections to~(\ref{2part}), which 
are due to multi holon/antiholon states and have a more complicated
structure, but can be shown to be important only at energies $\omega
\gg \Delta$~\cite{CET01} where the field-theory approach becomes invalid anyway.

The more general case with a relevant holon-antiholon interaction,
parameterized by $1/2 < \beta^{2} =K/2 < 1$, is discussed in
detail in Ref.~\cite{Iucci2006}; here, the parameter $K$ 
in a Luttinger-model description or the parameter $\beta^2$
in the sine-Gordon model characterize the strength of the
holon-antiholon interaction.
As we shall see below, our DDMRG is not sufficient to determine the value 
$\beta^2$ for the first Mott lobe accurately but we shall see that the assumption $\beta^2=1$ as used in Eq.~(15) 
provides a reasonable description of the DDMRG data for the frequency dependence of the current-current correlation 
function close to the gap, see below.

The current correlation function is finite only beyond the
single-particle gap with a square-root onset at $\omega=\Delta$.
It shows a maximum at $\omega\approx 1.24\Delta$ and a power-law decay at
large frequencies. 
The normalization $C_2(\Delta)$ cannot be calculated analytically.
To estimate the normalization $C_2(\Delta)$, we use the sum rule
for the conductivity,
\begin{equation}
\int_{0}^{\infty}\frac{{\rm d}\omega}{\pi} \re\left[\sigma(\omega)\right] = 
-\frac{1}{2L} \langle \Psi_0 | \hat{T} |\Psi_0 \rangle \; .
\label{eq:sumr}
\end{equation}
This quantity can be determined accurately using the ground-state DMRG.
For example, for $x=0.20$ we find $[\Delta=0.436t,
-\langle \Psi_0 | \hat{T} |\Psi_0 \rangle/(2L) = 1.418 t]$
and for $x=0.15$ we find
$[\Delta=1.63t, -\langle \Psi_0 | \hat{T} |\Psi_0 \rangle/(2L) = 1.133t]$.

Close to the Mott transition, most of the optical weight 
is concentrated at low energy $\omega \sim \Delta$.
Therefore, in the field-theory limit,
the l.h.s.~of~(\ref{eq:sumr}) can be determined 
from~(\ref{2part}) as the omitted terms have a negligible contribution
at low energy. This comparison provides $C_2(\Delta)$ for given $x$,
\begin{equation}
C_2(\Delta)=-\frac{\langle \Psi_0 | \hat{T} | \Psi_0 \rangle}{2 L} \frac{\pi}{F \Delta}
\end{equation}
with
\begin{eqnarray}
F&=& \frac{2}{\pi} \int_1^{\infty} {\rm d}y \frac{\sqrt{y^2-1}}{y^2} 
\nonumber \\
&&
\exp\left(-\int_0^\infty\frac{{\rm d}x}{x}\frac{1-\cos(2x{\rm Arccosh}(y)/\pi)\cosh x
}{
\exp(x/2)\cosh(x/2)\sinh x}\right) \nonumber \\
&\approx & 1.70[1] \; .
\end{eqnarray}
Therefore, for $x=0.20$, we have $C_2=6.01$ and,
for $x=0.15$, we find $C_2=1.28$.

The result from field theory is applicable in the region where the
gap is (exponentially) small. Therefore, it is complementary
to the DDMRG with its energy resolution $\eta\approx 12 W/L$
where $W=4t$ is the bare bandwidth.
We note that the formula~(\ref{2part}) works surprisingly
well for the fermionic Hubbard model even when 
the gap is not exponentially small but of the order of~$t$~\cite{JGE00,EGJ01}.
Therefore, we compare our DDMRG results for the current correlation function
for $x=0.20$ and $x=0.15$ 
to those from field theory in Sect.~\ref{subsec:currentcf}.

\section{Correlation functions}
\label{sec:cfs}

\subsection{Strong-coupling result to leading order}
\label{subsec:strongcoupling}

To leading order in the SC expansion,
the doubly occupied site and the hole only experience
their hard-core repulsion. Otherwise, they move freely with 
hoping amplitudes $t_h=t$ for the hole and $t_{d}=-2t$ for 
the double occupancy. Consequently,
both the current correlation function and the kinetic-energy
correlation function are given by a semi-ellipse,
\begin{equation}
\label{eq:sj_limit}
 S_{J,T}^{(0)}(\omega) =  \frac{4}{3 \pi} \sqrt{1-\left(\frac{\omega-U}{6 t}\right)^2} 
\; \Theta\left(1-\left(\frac{\omega-U}{6t}\right)^2\right) \;,
\end{equation}
where $\Theta(x)$ denotes the Heaviside-step-function, see also
Refs.~\cite{Ejima2011,TG11,Ejima2012}.
The result is qualitatively the same for the fermionic
Hubbard model~\cite{JGE00,EGJ01} where $t_{d}^{\rm F}=-t$.

The correlation function is finite only above the single-particle gap,
$\Delta\approx U-6t$ for $x\to 0$. Moreover, the correlation function
displays a square-root onset, as seen in field theory. Therefore,
the square-root onset above the single-particle gap
apparently is a generic feature of the correlation functions in the Mott-insulating phase.

\subsection{Current correlation function}
\label{subsec:currentcf}

\begin{figure}[ht]
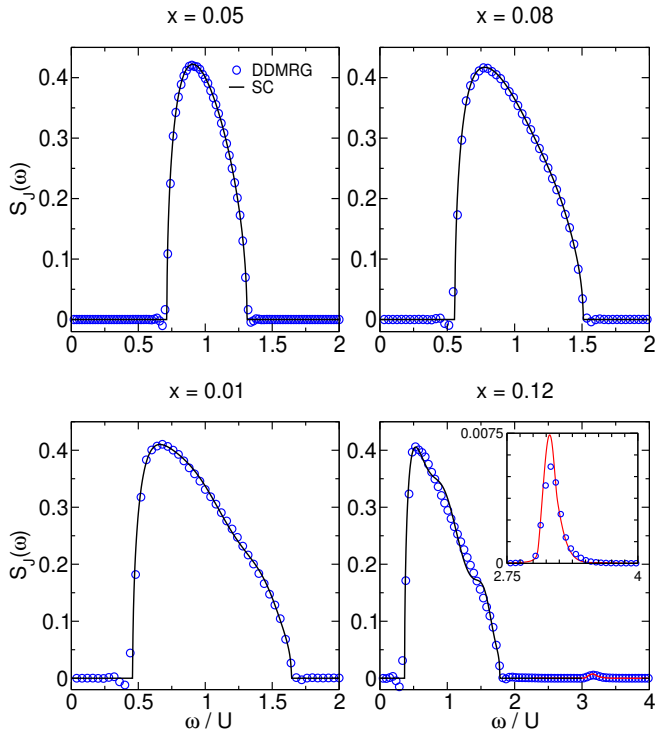

 \subfloat{\includegraphics[clip,width=0.25\textwidth, height=0.258\textwidth]{jj-x0.05n0.8dec.eps}}
 \subfloat{\includegraphics[clip,width=0.23\textwidth, height=0.258\textwidth]{jj-x0.08n0.8dec.eps}}  \\
 \subfloat{\includegraphics[clip,width=0.25\textwidth, height=0.258\textwidth]{jj-x0.1n0.8dec.eps}}
 \subfloat{\includegraphics[clip,width=0.23\textwidth, 
height=0.258\textwidth]{with-inset-jj-x0.12n0.8dec.eps}}
\caption{(Color online) Current correlation function for interactions $x=t/U=0.05,0.08,0.1,0.12$ 
from top left to bottom right. The deconvolved DDMRG data (circles) were obtained for $L=64$, PBC and $\eta=0.8t$.
Black solid lines give the result of the SC theory of the primary 
Hubbard band. Red solid lines give the result of the SC theory 
for the tertiary Hubbard band. A magnification of 
the tertiary Hubbard band is shown in the inset.\label{fig:jj_ddmrg_sc}}
\end{figure}

In Fig.~\ref{fig:jj_ddmrg_sc} 
we show the deconvolved DDMRG data and the results from sixth-order SC theory 
for the current correlation function
for $x=0.05$, $0.08$, $0.10$, and $0.12$. 
For $x=0.05$, the correlation function is almost an undisturbed
semi-ellipse, see Eq.~(\ref{eq:sj_limit}).
On decreasing the interaction, 
the width of the primary Hubbard band increases, and spectral weight is shifted
towards the lower band edge so that 
the current correlation function becomes asymmetric.
As seen from the figure, both DDMRG and SC perfectly agree with each other
up to $x=t/U=0.08$. As discussed in Sect.~\ref{subsec:broadening},
the slightly negative DDMRG data for the current correlation function
close to the band edges are an artifact of the deconvolution scheme.

For smaller interactions, $x\geq 0.10$, the current correlation function steeply rises to its 
maximum value as a function of frequency, and falls off monotonically 
beyond the maximum.
The agreement of DDMRG and SC is satisfactory only at first glance
because some wiggles appear in the SC results that are absent in the DDMRG data.
The wiggles are more pronounced for $x=0.12$ than for $x=0.10$.
The reason as to why these unphysical wiggles appear in SC
is discussed in Appendix~\ref{app:dev}.

For $x=0.12$,  the tertiary Hubbard band can be seen in the DDMRG data. 
In the SC theory this band emerges 
from states with a triple occupancy and two holes whose contributions 
to $|\tilde{J}|\Phi_0\rangle|^2$ are of the order 
$\mathcal{O}(x^4)$.
For $x=0.12$ the weight of the primary, secondary, and tertiary Hubbard bands is 
given by
$3.04$, $1.5\times 10^{-3}$, and $0.01$, respectively. 
Here, the weight of the primary Hubbard band has been 
approximated by a sixth order expansion, while a third and fourth order 
expansion has been used for the secondary and tertiary 
Hubbard band, respectively.
The secondary Hubbard band is described by two double occupancies and 
two holes but is not visible because its weight is only of the order $\mathcal{O}(x^6)$, see 
Appendix~\ref{app:diagrams}.

To leading order, the total weight of the correlation functions
is given by $\int_0^{\infty} S_{T,J}^{(0)}(\omega) \rmd \omega= 4$.
For $x>0$, the total spectral weight can be determined by the static DMRG 
method with great precision so that it
can be used as a test for the accuracy of the SC expansion.
The results for the total weight from DMRG and SC deviate
by less than  $3 \times 10^{-3}$ up to $x \leq 0.12$.
For higher values of $x$, the SC theory seems to overestimate the weight.

\begin{figure}[ht!]
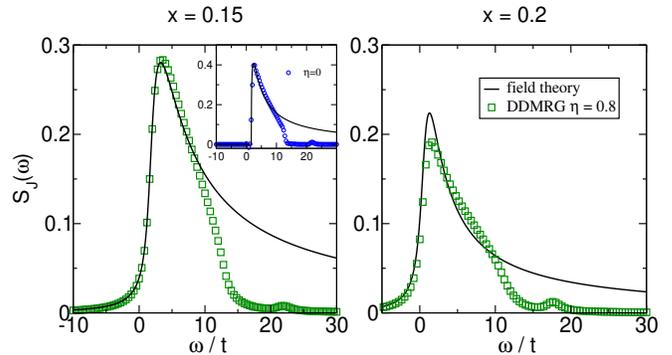

 \subfloat{\includegraphics[clip,width=0.25\textwidth, height=0.258\textwidth]{jj-x0.15n0.8.eps}}
 \subfloat{\includegraphics[clip,width=0.23\textwidth, height=0.258\textwidth]{jj-x0.2n0.8.eps}}
 \caption{(Color online) DDRMG data (squares) for $L=64$, PBC and $\eta=0.8t$ 
for $x=0.15$ (left) and $x=0.2$ (right). 
The solid black line gives the Lorentz-broadened field-theoretical 
results. The deconvolved DDMRG data (circles) and the field-theoretical
results are shown in the inset for $x=0.15$.\label{fig:jj_ddmrg_large_x}}
\end{figure}

For interaction parameters $x=0.15$ and $x=0.20$, only the 
DDMRG can provide reliable results. 
In Fig.~\ref{fig:jj_ddmrg_large_x} we plot
the Lorentz-broadened current correlation function
$S_J^{\eta}(\omega)$ for $\eta=0.8t$ and compare it to the predictions from field theory.
Even up to half the width of the primary Hubbard band,
the field-theory curve agrees with the DDMRG data, 
despite the fact that the gap is quite sizable for $x=0.15$.
For $x=0.20$, the agreement appears to be worse than for $x=0.15$.
At $x=0.20$, however, the DDMRG has difficulties to resolve the quite sharp 
maximum close to the onset of the correlation function.
Here, larger system sizes and, correspondingly, smaller values for
$\eta$ would improve the agreement.
Therefore, we show exemplarily a comparison of the deconvolved DDMRG data with the field theory results 
in the inset of Fig.~\ref{fig:jj_ddmrg_large_x} for $x=0.15$ and omit a further deconvolution of the DDMRG data
in this parameter regime.

\subsection{Kinetic correlation function}
\label{sec:kincf}

Before we discuss $S_T(\omega>0)$, we briefly comment on the 
weight at $\omega=0$. Within the SC expansion to sixth order,
it is given by
\begin{equation}
\sqrt{w_0^T} = 8x-16x^3+\frac{529}{3}x^5+\mathcal{O}(x^7) \;.
\end{equation}
 Apparently, the series converges rapidly even for $x=0.2$.
In general, the SC expansion 
reproduces the total weight of the kinetic-energy correlation function
with an accuracy of $3.7 \times 10^{-2}$ for $x \leq 0.12$.
As for the current correlation function, the SC expansion 
seems to overestimate the weight  for higher values of $x$.

\begin{figure}[ht]
 \subfloat{\includegraphics[clip, width=0.25\textwidth, height=0.258\textwidth]
 {tt-x0.05n0.8dec.eps}}
 \subfloat{\includegraphics[clip,width=0.23\textwidth, height=0.258\textwidth]
 {tt-x0.08n0.8dec.eps}} \\
\subfloat{\includegraphics[clip,width=0.25\textwidth, 
height=0.258\textwidth]{with-inset-tt-x0.1n0.8dec.eps}}
  \subfloat{\includegraphics[clip,width=0.23\textwidth, 
height=0.258\textwidth]{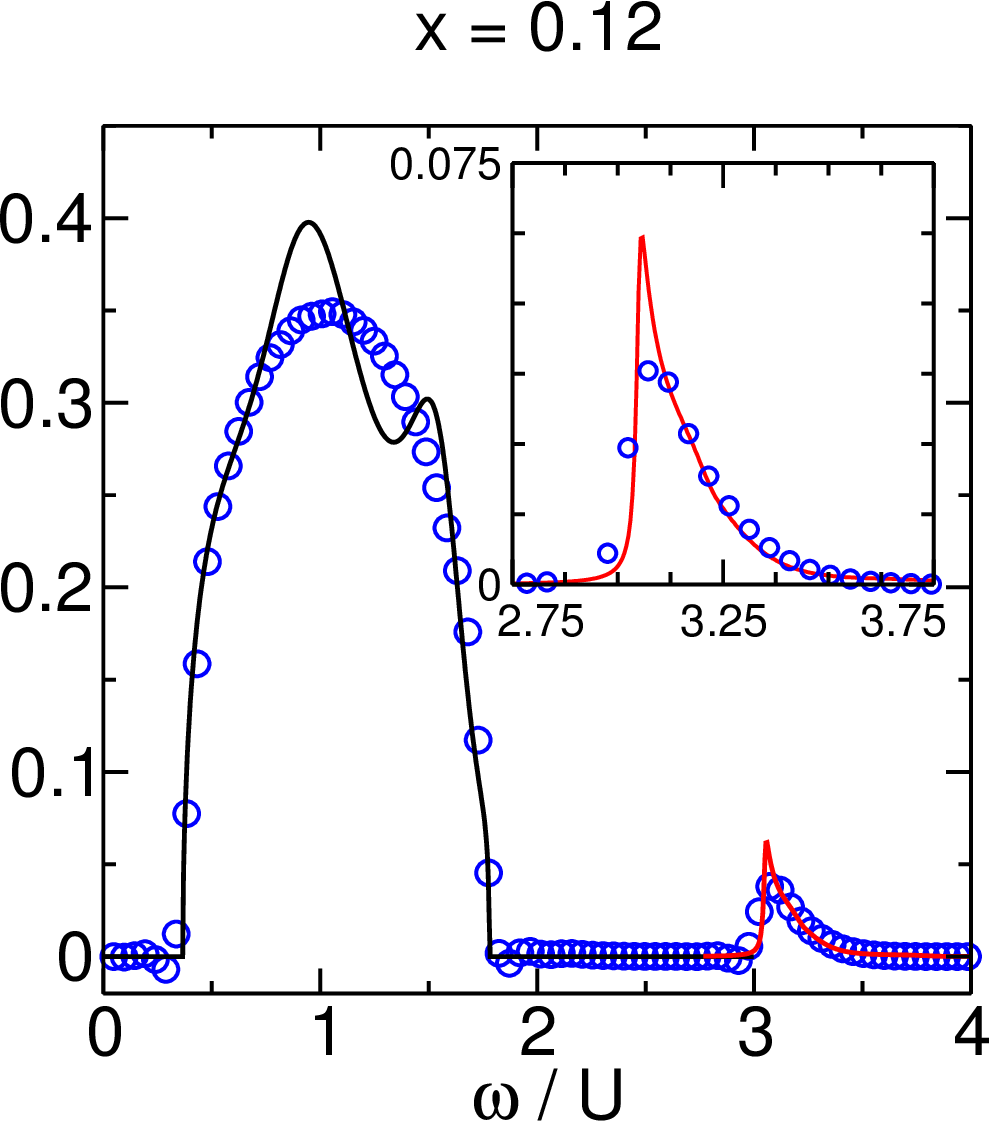}}
\caption{(Color online) Kinetic-energy correlation function for interactions
$x=0.05,0.08,0.1,0.12$ (from top left to bottom right). 
The deconvolved DDMRG data (circles) were obtained for $L=64$, PBC and $\eta=0.8t$. 
Black solid lines gives the result of the SC theory of the primary Hubbard band. 
Red solid line gives the result of the SC theory for the 
tertiary Hubbard band. A magnification of the tertiary Hubbard band 
is shown in the insets.\label{fig:tt_ddmrg_sc}}
\end{figure}

In Figure~\ref{fig:tt_ddmrg_sc} we show 
the kinetic-energy correlation function
for the values $x=0.05$, $0.08$, $0.1$, and $0.12$. 
For $x\leq 0.08$ the agreement of SC and DDMRG is very good.
For $x=0.1$ and $0.12$, the sixth-order SC theory starts to develop
wiggles as in the case of the current correlation functions,
which are not seen in the DDMRG data. Therefore, the SC expansion
cannot be used beyond $x=0.10$.
As the current correlation function, the kinetic-energy correlation function
is finite above the single-particle gap with a square-root onset.
In contrast to the current correlation function, the 
primary Hubbard band for $S_T$ appears to remain symmetrical
so that its maximum appears in the middle of the band for all
interaction strengths.

The total weight of the secondary and tertiary Hubbard bands are of the order 
$\mathcal{O}(x^6)$ and $\mathcal{O}(x^2)$. 
For $x=0.12$, the weight of the primary, secondary, and tertiary Hubbard 
bands is given by $3.3$, $3.3\times 10^{-3}$, and $0.076$, respectively.
The operator $\hat{T}$ is symmetric under spatial inversion, so 
that also states in which the holes are symmetrically placed to the left and 
the right of the triple occupancy occur in $\tilde{T}|\Phi_0\rangle$. They are 
the leading-order contributions to the tertiary Hubbard band.
Therefore, the tertiary Hubbard band is more prominent in $S_T$ than in $S_J$ 
where the spatial antisymmetry eliminates these states, 
see Appendix~\ref{app:diagrams}.

\begin{figure}[ht]
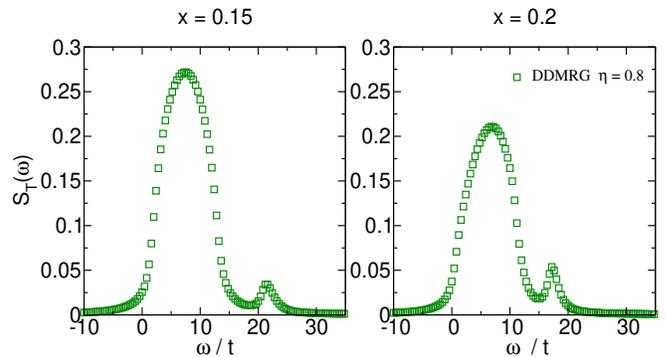

 \subfloat{\includegraphics[clip,width=0.25\textwidth, height=0.258\textwidth]{tt-x0.15n0.8.eps}}
 \subfloat{\includegraphics[clip,width=0.23\textwidth, height=0.258\textwidth]{tt-x0.2n0.8.eps}}
 \caption{(Color online) DDRMG data (squares) were obtained for $L=64$, PBC and $\eta=0.8t$ 
for $x=0.15$ (left) and $x=0.2$ (right).\label{fig:tt_ddmrg_large_x}}
\end{figure}

In Fig.~\ref{fig:tt_ddmrg_sc},
we also compare the DDMRG data with SC results for the tertiary Hubbard band.
Note, however, that the calculations are quite cumbersome so that
we used the weights of the tertiary Hubbard band up to fourth
order and considered the action of the effective Hamiltonian on $\mathcal{H}_3$ 
in second order only. This means that we work with $\hat{S}_4$ and $\hat{h}_1$.  Given these simplifications,
we obtain the SC results for the tertiary Hubbard band from an exact 
diagonalization of a system of $L=128$ lattice sites and 
a very small spectral broadening of $\eta=0.08$. 
Despite these limitations, 
our restricted SC approach quite accurately describes the 
asymmetrical shape of the tertiary Hubbard band. 

The third 
Hubbard band shows some interesting 
features. For example, an attractive force acts between the two 
particles if they are placed next to each other. Moreover, the alignment of a hole next to 
the triple occupancy leads to a decrease in the potential energy. These effects induce
sizable correlations between the holes and the triple occupancy which influence  
 the shape of the correlation function.

For interaction parameters $x=0.15$ and $x=0.20$, only the 
DDMRG can provide reliable results. 
In Fig.~\ref{fig:tt_ddmrg_large_x} we plot the Lorentz-broadened
kinetic-energy correlation function for $\eta=0.8t$.
The weight of the primary Hubbard band shrinks when we approach the transition
whereas the tertiary Hubbard band gains more weight as a function of~$x$.
Therefore, the tertiary Hubbard band becomes clearly visible
for $x=0.20$.

\section{Conclusion and Outlook}
\label{sec:outlook}

In this work we calculated the current and kinetic-energy correlation functions
for the Mott insulating regime of the one-dimensional Bose--Hubbard model at 
filling $\rho=N/L=1$ using the strong-coupling (SC) expansion up to sixth order
in $x=t/U$ 
and the dynamical density-matrix renormalization group (DDMRG) method
on rings with $L=64$ lattice sites.
The DDMRG data for finite $\eta=0.8t$ permit a reliable deconvolution
of the Lorentz-broadened data so that the correlation functions can be studied 
in the thermodynamical limit.

A comparison of sixth-order SC and DDMRG results shows that SC is reliable
up to $x=0.10$. DDMRG on $L=64$ sites can be used 
up to $x\lesssim 0.20$. For $x\gtrsim 0.20$ it becomes difficult to resolve
the sharp maximum in the current correlation functions at 
frequency $\omega \approx 1.3\Delta$ where $\Delta$ is the 
single-particle gap.
In any case, the exponentially small gap close to the Mott transition
at $x_{\rm c}\approx 0.305$ cannot be resolved by (D)DMRG
for $x\gtrsim 0.25$. 

The correlation functions are dominated by the
primary Hubbard band around $\omega \approx U$. 
The primary Hubbard band starts at the single-particle gap
with a characteristic square-root onset 
$S_{J,T}(\omega \to \Delta) \propto \sqrt{\omega-\Delta}$.
This is seen in SC perturbation theory and in the deconvolved
DDMRG data. It is confirmed for the current correlation function
by field theory which is applicable close to the Mott transition.
Apart from a maximum at low frequencies,
the primary Hubbard band of the current correlation function is featureless.
The primary Hubbard band of the kinetic-energy correlation function
appears to be symmetric with a single maximum at the band center.
For both correlation functions, the secondary
Hubbard band is very small but the tertiary Hubbard band around $\omega=3U$
becomes visible for $x\gtrsim 0.10$.
The  asymmetric shape is understood from SC theory which includes correlations between the 
quasiparticles.

For the primary Hubbard band in the one-dimensional Bose--Hubbard
model, the attractive correlations between doubly occupied site and hole
are not significant enough to overcome their hard-core repulsion.
Therefore, it requires a finite nearest-neighbor interaction
to generate an excitonic state~\cite{KWM00,JGE00,EGJ01}.
The situation could change 
in the two-dimensional Bose--Hubbard model. In recent quantum Monte-Carlo studies
and field-theoretical studies~\cite{Pollet2012},
the existence of a peak in the optical conductivity
has been alluded in the critical region above but close to the Mott transition, 
$x_{\rm c}^{2d} -x  < 1.15 x_{\rm c}^{2d}$.
This peak is a signature of the Higgs boson in the two-dimensional
superfluid which should persist as a resonance in the Mott phase.

Albeit the SC expansion is better behaved in two than in one
dimension~\cite{Teichmann2009a}, it is not clear at this point
whether or not the expansion can be carried out far enough to include such features. The SC diagrams
show that higher orders in the expansion generate
an attractive interaction between the quasiparticles in the primary and tertiary Hubbard band.
However, it is unclear whether or not these 
interactions
are sufficient to generate a resonance structure close to the transition.
Moreover, the diagrammatic expansion is considerably more cumbersome
in two dimension than in one dimension. Therefore, 
a SC study of the two-dimensional Bose--Hubbard model
remains an open problem.

\acknowledgments
S.E. and H.F. gratefully acknowledge financial support 
by the Deutsche Forschungsgemeinschaft through SFB 652.

\appendix

\section{Diagrammatic analysis}
\label{app:diagrams}

The Taylor expansion for $\hat{h}$ in Eq.~(\ref{eq:harris_lange_transformation})
gives rise to so-called 
{\sl process chains}~\cite{Eckardt2009,Damski2006,Teichmann2009a}. 
They represent sequences that involve alternately
the kinetic-energy operator~$\hat{T}$ and projection operators 
$\hat{P}_D$ onto subspaces $\mathcal{H}_D$. 
For example, the leading-order approximations of the 
transformation operator $\hat{S}$ 
and the effective Hamilton operator $\hat{h}$ can be written as
\begin{eqnarray}
U\hat{D} + t \hat{h}_0 &=& U\hat{D}+ t\sum_{D}\hat{P}_D \hat{T} \hat{P}_{D} \; , \\
\hat{S}_1 &=& \sum_{D_1,D_2} \frac{\hat{P}_{D_1} \hat{T} \hat{P}_{D_2}}{D_1-D_2}\;.
\end{eqnarray}
This example shows that each process chain 
must be weighted according to the number of double 
occupancies that occur in intermediate steps.

The application of the Baker-Campbell-Hausdorff 
formula to eqs.~(\ref{eq:harris_lange_transformation}) shows that 
$\hat{h}$ and $\hat{S}$ can be written in terms of nested Lie-brackets. 
In this way, only connected hopping processes 
must be considered in the action of these operators onto some state. This simplification 
leads to an intuitive understanding as to 
why the secondary Hubbard band has vanishingly small intensity.
Its starting state consists of two double occupancies and two holes. 
The leading order contribution for the secondary Hubbard band can be depicted as
\begin{figure}[h]
 \subfloat{\includegraphics[clip,width=0.12\textwidth]{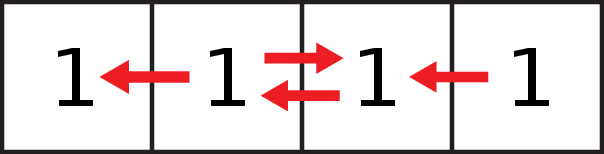}}\;,
 \subfloat{\includegraphics[clip,width=0.12\textwidth]{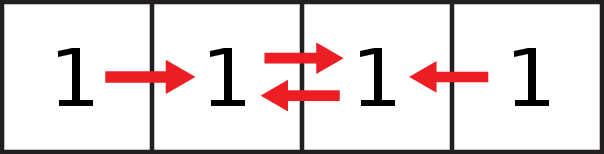}}\;,
  \subfloat{\includegraphics[clip,width=0.12\textwidth]{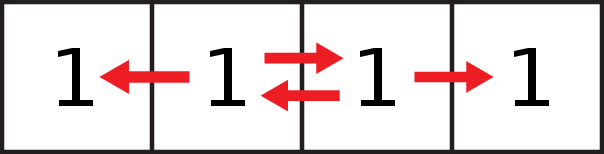}
\label{fig:diagrams}}
\quad . \hfill
\end{figure}

\noindent The actual computation of these diagrams requires 
the evaluation of all possible hopping processes generated by an 
arbitrary time ordering of its constituents.
One hopping process originates from the operator $\hat{T}$ or $\hat{J}$ 
itself while the other processes result from $\hat{S}_3$ 
so that these states are of the order $\mathcal{O}(x^3)$. 
Note that the antisymmetry of $\hat{J}$ with respect to spatial inversion leads to a 
cancellation of the last two diagrams.

\section{Limitations of the strong-coupling expansion}
\label{app:dev}

In the primary Hubbard band, 
a particle hole-pair is created in which the hole is placed $n$ 
lattice sites to the right or left of the double occupancy (state $|\pm n\rangle$).
In the lowest-order approximation, the hole 
cannot jump over the double occupancy, so that it moves on an open chain 
on the remaining $L-1$ sites.
The correlation functions are a weighted sum 
of Green functions of the form $G_{ij}(\omega) = 
\langle i|  \delta(\omega - (\hat{h}-E_0)) | j \rangle $. 

For $x \geq 0.1$, the SC expansion 
shows some wiggles which can be understood as follows. 
In general, the Greens functions $G_{jj}$ or $G_{1j}$ 
of a particle in an open chain subject to nearest-neighbor 
hopping with lattice site index $j$  will have $j$ extrema. 
The parts of the state $ \tilde{J}  |\Phi_0 \rangle$ 
that describe particle-hole pairs which are separated by $j$ lattice sites thus lead to 
contributions in $S_{J}(\omega)$ which have $j$ maxima. The wiggles
naturally appear when the expansion parameter~$x$ is chosen too large so 
that the weight of these contributions is overestimated and not yet corrected by 
contributions to higher orders.


\begin{thebibliography}{27}%
\makeatletter
\providecommand \@ifxundefined [1]{%
 \@ifx{#1\undefined}
}%
\providecommand \@ifnum [1]{%
 \ifnum #1\expandafter \@firstoftwo
 \else \expandafter \@secondoftwo
 \fi
}%
\providecommand \@ifx [1]{%
 \ifx #1\expandafter \@firstoftwo
 \else \expandafter \@secondoftwo
 \fi
}%
\providecommand \natexlab [1]{#1}%
\providecommand \enquote  [1]{``#1''}%
\providecommand \bibnamefont  [1]{#1}%
\providecommand \bibfnamefont [1]{#1}%
\providecommand \citenamefont [1]{#1}%
\providecommand \href@noop [0]{\@secondoftwo}%
\providecommand \href [0]{\begingroup \@sanitize@url \@href}%
\providecommand \@href[1]{\@@startlink{#1}\@@href}%
\providecommand \@@href[1]{\endgroup#1\@@endlink}%
\providecommand \@sanitize@url [0]{\catcode `\\12\catcode `\$12\catcode
  `\&12\catcode `\#12\catcode `\^12\catcode `\_12\catcode `\%12\relax}%
\providecommand \@@startlink[1]{}%
\providecommand \@@endlink[0]{}%
\providecommand \url  [0]{\begingroup\@sanitize@url \@url }%
\providecommand \@url [1]{\endgroup\@href {#1}{\urlprefix }}%
\providecommand \urlprefix  [0]{URL }%
\providecommand \Eprint [0]{\href }%
\providecommand \doibase [0]{http://dx.doi.org/}%
\providecommand \selectlanguage [0]{\@gobble}%
\providecommand \bibinfo  [0]{\@secondoftwo}%
\providecommand \bibfield  [0]{\@secondoftwo}%
\providecommand \translation [1]{[#1]}%
\providecommand \BibitemOpen [0]{}%
\providecommand \bibitemStop [0]{}%
\providecommand \bibitemNoStop [0]{.\EOS\space}%
\providecommand \EOS [0]{\spacefactor3000\relax}%
\providecommand \BibitemShut  [1]{\csname bibitem#1\endcsname}%
\let\auto@bib@innerbib\@empty
\bibitem [{\citenamefont {Bloch}\ \emph {et~al.}(2008)\citenamefont {Bloch},
  \citenamefont {Dalibard},\ and\ \citenamefont {Zwerger}}]{RevModPhys.80.885}%
  \BibitemOpen
  \bibfield  {author} {\bibinfo {author} {\bibfnamefont {I.}~\bibnamefont
  {Bloch}}, \bibinfo {author} {\bibfnamefont {J.}~\bibnamefont {Dalibard}}, \
  and\ \bibinfo {author} {\bibfnamefont {W.}~\bibnamefont {Zwerger}},\ }\href
  {\doibase 10.1103/RevModPhys.80.885} {\bibfield  {journal} {\bibinfo
  {journal} {Rev. Mod. Phys.}\ }\textbf {\bibinfo {volume} {80}},\ \bibinfo
  {pages} {885} (\bibinfo {year} {2008})}\BibitemShut {NoStop}%
\bibitem [{\citenamefont {Dalibard}\ \emph {et~al.}(2011)\citenamefont
  {Dalibard}, \citenamefont {Gerbier}, \citenamefont
  {Juzeli\ifmmode~\bar{u}\else \={u}\fi{}nas},\ and\ \citenamefont
  {\"Ohberg}}]{RevModPhys.83.1523}%
  \BibitemOpen
  \bibfield  {author} {\bibinfo {author} {\bibfnamefont {J.}~\bibnamefont
  {Dalibard}}, \bibinfo {author} {\bibfnamefont {F.}~\bibnamefont {Gerbier}},
  \bibinfo {author} {\bibfnamefont {G.}~\bibnamefont
  {Juzeli\ifmmode~\bar{u}\else \={u}\fi{}nas}}, \ and\ \bibinfo {author}
  {\bibfnamefont {P.}~\bibnamefont {\"Ohberg}},\ }\href {\doibase
  10.1103/RevModPhys.83.1523} {\bibfield  {journal} {\bibinfo  {journal} {Rev.
  Mod. Phys.}\ }\textbf {\bibinfo {volume} {83}},\ \bibinfo {pages} {1523}
  (\bibinfo {year} {2011})}\BibitemShut {NoStop}%
\bibitem [{\citenamefont {Struck}\ \emph {et~al.}(2012)\citenamefont {Struck},
  \citenamefont {\"Olschl\"ager}, \citenamefont {Weinberg}, \citenamefont
  {Hauke}, \citenamefont {Simonet}, \citenamefont {Eckardt}, \citenamefont
  {Lewenstein}, \citenamefont {Sengstock},\ and\ \citenamefont
  {Windpassinger}}]{PhysRevLett.108.225304}%
  \BibitemOpen
  \bibfield  {author} {\bibinfo {author} {\bibfnamefont {J.}~\bibnamefont
  {Struck}}, \bibinfo {author} {\bibfnamefont {C.}~\bibnamefont
  {\"Olschl\"ager}}, \bibinfo {author} {\bibfnamefont {M.}~\bibnamefont
  {Weinberg}}, \bibinfo {author} {\bibfnamefont {P.}~\bibnamefont {Hauke}},
  \bibinfo {author} {\bibfnamefont {J.}~\bibnamefont {Simonet}}, \bibinfo
  {author} {\bibfnamefont {A.}~\bibnamefont {Eckardt}}, \bibinfo {author}
  {\bibfnamefont {M.}~\bibnamefont {Lewenstein}}, \bibinfo {author}
  {\bibfnamefont {K.}~\bibnamefont {Sengstock}}, \ and\ \bibinfo {author}
  {\bibfnamefont {P.}~\bibnamefont {Windpassinger}},\ }\href {\doibase
  10.1103/PhysRevLett.108.225304} {\bibfield  {journal} {\bibinfo  {journal}
  {Phys. Rev. Lett.}\ }\textbf {\bibinfo {volume} {108}},\ \bibinfo {pages}
  {225304} (\bibinfo {year} {2012})}\BibitemShut {NoStop}%
\bibitem [{\citenamefont {K{\"u}hner}\ \emph {et~al.}(2000)\citenamefont
  {K{\"u}hner}, \citenamefont {White},\ and\ \citenamefont {Monien}}]{KWM00}%
  \BibitemOpen
  \bibfield  {author} {\bibinfo {author} {\bibfnamefont {T.~D.}\ \bibnamefont
  {K{\"u}hner}}, \bibinfo {author} {\bibfnamefont {S.~R.}\ \bibnamefont
  {White}}, \ and\ \bibinfo {author} {\bibfnamefont {H.}~\bibnamefont
  {Monien}},\ }\href@noop {} {\bibfield  {journal} {\bibinfo  {journal} {Phys.
  Rev. B}\ }\textbf {\bibinfo {volume} {61}},\ \bibinfo {pages} {12474}
  (\bibinfo {year} {2000})}\BibitemShut {NoStop}%
\bibitem [{\citenamefont {Huo}\ \emph {et~al.}(2011)\citenamefont {Huo},
  \citenamefont {Zhang}, \citenamefont {Chen}, \citenamefont {Troyer},\ and\
  \citenamefont {Schollw\"ock}}]{PhysRevA.84.043608}%
  \BibitemOpen
  \bibfield  {author} {\bibinfo {author} {\bibfnamefont {J.-W.}\ \bibnamefont
  {Huo}}, \bibinfo {author} {\bibfnamefont {F.-C.}\ \bibnamefont {Zhang}},
  \bibinfo {author} {\bibfnamefont {W.}~\bibnamefont {Chen}}, \bibinfo {author}
  {\bibfnamefont {M.}~\bibnamefont {Troyer}}, \ and\ \bibinfo {author}
  {\bibfnamefont {U.}~\bibnamefont {Schollw\"ock}},\ }\href {\doibase
  10.1103/PhysRevA.84.043608} {\bibfield  {journal} {\bibinfo  {journal} {Phys.
  Rev. A}\ }\textbf {\bibinfo {volume} {84}},\ \bibinfo {pages} {043608}
  (\bibinfo {year} {2011})}\BibitemShut {NoStop}%
\bibitem [{\citenamefont {Ejima}\ \emph {et~al.}(2011)\citenamefont {Ejima},
  \citenamefont {Fehske},\ and\ \citenamefont {Gebhard}}]{Ejima2011}%
  \BibitemOpen
  \bibfield  {author} {\bibinfo {author} {\bibfnamefont {S.}~\bibnamefont
  {Ejima}}, \bibinfo {author} {\bibfnamefont {H.}~\bibnamefont {Fehske}}, \
  and\ \bibinfo {author} {\bibfnamefont {F.}~\bibnamefont {Gebhard}},\ }\href
  {\doibase 10.1209/0295-5075/93/30002} {\bibfield  {journal} {\bibinfo
  {journal} {Europhys.\ Lett.}\ }\textbf {\bibinfo {volume} {93}},\ \bibinfo
  {pages} {30002} (\bibinfo {year} {2011})}\BibitemShut {NoStop}%
\bibitem [{\citenamefont {Tokuno}\ and\ \citenamefont
  {Giamarchi}(2011)}]{TG11}%
  \BibitemOpen
  \bibfield  {author} {\bibinfo {author} {\bibfnamefont {A.}~\bibnamefont
  {Tokuno}}\ and\ \bibinfo {author} {\bibfnamefont {T.}~\bibnamefont
  {Giamarchi}},\ }\href@noop {} {\bibfield  {journal} {\bibinfo  {journal}
  {Phys. Rev. Lett.}\ }\textbf {\bibinfo {volume} {106}},\ \bibinfo {pages}
  {205301} (\bibinfo {year} {2011})};\ \bibinfo {note} {the analytic expression
  for the leading order of the correlation functions can already be extracted
  from the results presented in this work. However, our band width seems to be
  twice as large, which might be caused by a wrong scaling of $t$. Note also
  the different values for $x_c$.}\BibitemShut {Stop}%
\bibitem [{\citenamefont {Iucci}\ \emph {et~al.}(2006)\citenamefont {Iucci},
  \citenamefont {Cazalilla}, \citenamefont {Ho},\ and\ \citenamefont
  {Giamarchi}}]{Iucci2006}%
  \BibitemOpen
  \bibfield  {author} {\bibinfo {author} {\bibfnamefont {A.}~\bibnamefont
  {Iucci}}, \bibinfo {author} {\bibfnamefont {M.A.}~\bibnamefont {Cazalilla}},
  \bibinfo {author} {\bibfnamefont {A.F.}~\bibnamefont {Ho}}, \ and\ \bibinfo
  {author} {\bibfnamefont {T.}~\bibnamefont {Giamarchi}},\ }\href {\doibase
  10.1103/PhysRevA.73.041608} {\bibfield  {journal} {\bibinfo  {journal} {Phys.
  Rev. A}\ }\textbf {\bibinfo {volume} {73}},\ \bibinfo {pages} {041608}
  (\bibinfo {year} {2006})}\BibitemShut {NoStop}%
\bibitem [{\citenamefont {Kollath}\ \emph {et~al.}(2006)\citenamefont
  {Kollath}, \citenamefont {Iucci}, \citenamefont {McCulloch},\ and\
  \citenamefont {Giamarchi}}]{Kollath2006}%
  \BibitemOpen
  \bibfield  {author} {\bibinfo {author} {\bibfnamefont {C.}~\bibnamefont
  {Kollath}}, \bibinfo {author} {\bibfnamefont {A.}~\bibnamefont {Iucci}},
  \bibinfo {author} {\bibfnamefont {I.~P.}\ \bibnamefont {McCulloch}}, \ and\
  \bibinfo {author} {\bibfnamefont {T.}~\bibnamefont {Giamarchi}},\ }\href
  {\doibase 10.1103/PhysRevA.74.041604} {\bibfield  {journal} {\bibinfo
  {journal} {Phys. Rev. A}\ }\textbf {\bibinfo {volume} {74}},\ \bibinfo
  {pages} {041604} (\bibinfo {year} {2006})}\BibitemShut {NoStop}%
\bibitem [{\citenamefont {Mahan}(2000)}]{mahan2000}%
  \BibitemOpen
  \bibfield  {author} {\bibinfo {author} {\bibfnamefont {G.~D.}\ \bibnamefont
  {Mahan}},\ }\href@noop {} {\emph {\bibinfo {title} {{Many-Particle
  Physics}}}}\ (\bibinfo  {publisher} {Springer},\ \bibinfo {year}
  {2000})\BibitemShut {NoStop}%
\bibitem [{\citenamefont {Ejima}\ \emph
  {et~al.}(2012{\natexlab{a}})\citenamefont {Ejima}, \citenamefont {Fehske},
  \citenamefont {Gebhard}, \citenamefont {zu~M\"{u}nster}, \citenamefont
  {Knap}, \citenamefont {Arrigoni},\ and\ \citenamefont {von~der
  Linden}}]{Ejima2012}%
  \BibitemOpen
  \bibfield  {author} {\bibinfo {author} {\bibfnamefont {S.}~\bibnamefont
  {Ejima}}, \bibinfo {author} {\bibfnamefont {H.}~\bibnamefont {Fehske}},
  \bibinfo {author} {\bibfnamefont {F.}~\bibnamefont {Gebhard}}, \bibinfo
  {author} {\bibfnamefont {K.}~\bibnamefont {zu~M\"{u}nster}}, \bibinfo
  {author} {\bibfnamefont {M.}~\bibnamefont {Knap}}, \bibinfo {author}
  {\bibfnamefont {E.}~\bibnamefont {Arrigoni}}, \ and\ \bibinfo {author}
  {\bibfnamefont {W.}~\bibnamefont {von~der Linden}},\ }\href {\doibase
  10.1103/PhysRevA.85.053644} {\bibfield  {journal} {\bibinfo  {journal} {Phys.
  Rev. A}\ }\textbf {\bibinfo {volume} {85}},\ \bibinfo {pages} {053644}
  (\bibinfo {year} {2012}{\natexlab{a}})}\BibitemShut {NoStop}%
\bibitem [{\citenamefont {Ejima}\ \emph
  {et~al.}(2012{\natexlab{b}})\citenamefont {Ejima}, \citenamefont {Fehske},\
  and\ \citenamefont {Gebhard}}]{EFG12}%
  \BibitemOpen
  \bibfield  {author} {\bibinfo {author} {\bibfnamefont {S.}~\bibnamefont
  {Ejima}}, \bibinfo {author} {\bibfnamefont {H.}~\bibnamefont {Fehske}}, \
  and\ \bibinfo {author} {\bibfnamefont {F.}~\bibnamefont {Gebhard}},\
  }\href@noop {} {\bibfield  {journal} {\bibinfo  {journal} {J.\ Phys.\ Conf.\
  Ser.}\ }\textbf {\bibinfo {volume} {391}},\ \bibinfo {pages} {012031}
  (\bibinfo {year} {2012}{\natexlab{b}})}\BibitemShut {NoStop}%
\bibitem [{\citenamefont {Ejima}\ \emph {et~al.}(2013)\citenamefont {Ejima},
  \citenamefont {Lange}, \citenamefont {Fehske}, \citenamefont {Gebhard},\ and\
  \citenamefont {zu~M\"{u}nster}}]{Ejima2013}%
  \BibitemOpen
  \bibfield  {author} {\bibinfo {author} {\bibfnamefont {S.}~\bibnamefont
  {Ejima}}, \bibinfo {author} {\bibfnamefont {F.}~\bibnamefont {Lange}},
  \bibinfo {author} {\bibfnamefont {H.}~\bibnamefont {Fehske}}, \bibinfo
  {author} {\bibfnamefont {F.}~\bibnamefont~{Gebhard}}, and\ \bibinfo
  {author} {\bibfnamefont {K.}~\bibnamefont {zu~M\"{u}nster}},\ }\href {\doibase
  10.1103/PhysRevA.88.063625} {\bibfield  {journal} {\bibinfo  {journal} {Phys.
  Rev. A}\ }\textbf {\bibinfo {volume} {88}},\ \bibinfo {pages} {063625}
  (\bibinfo {year} {2013})}\BibitemShut {NoStop}%
\bibitem [{\citenamefont {K\"{u}hner}\ and\ \citenamefont
  {White}(1999)}]{KW99}%
  \BibitemOpen
  \bibfield  {author} {\bibinfo {author} {\bibfnamefont {T.D.}~\bibnamefont
  {K\"{u}hner}}\ and\ \bibinfo {author} {\bibfnamefont {S.R.}~\bibnamefont
  {White}},\ }\href {\doibase 10.1103/PhysRevB.60.335} {\bibfield  {journal}
  {\bibinfo  {journal} {Phys. Rev. B}\ }\textbf {\bibinfo {volume} {60}},\
  \bibinfo {pages} {335} (\bibinfo {year} {1999})}\BibitemShut {NoStop}%
\bibitem [{\citenamefont {White}(1992)}]{Wh92}%
  \BibitemOpen
  \bibfield  {author} {\bibinfo {author} {\bibfnamefont {S.R.}\ \bibnamefont
  {White}},\ }\href@noop {} {\bibfield  {journal} {\bibinfo  {journal} {Phys.
  Rev. Lett.}\ }\textbf {\bibinfo {volume} {69}},\ \bibinfo {pages} {2863}
  (\bibinfo {year} {1992})}\BibitemShut {NoStop}%
\bibitem [{\citenamefont {Jeckelmann}(2002)}]{Je02b}%
  \BibitemOpen
  \bibfield  {author} {\bibinfo {author} {\bibfnamefont {E.}~\bibnamefont
  {Jeckelmann}},\ }\href {\doibase 10.1103/PhysRevB.66.045114} {\bibfield
  {journal} {\bibinfo  {journal} {Phys. Rev. B}\ }\textbf {\bibinfo {volume}
  {66}},\ \bibinfo {pages} {045114} (\bibinfo {year} {2002})}\BibitemShut
  {NoStop}%
  \bibitem [{\citenamefont {Jeckelmann}\ and\ \citenamefont
  {Fehske}(2007)}]{JF07}%
  \BibitemOpen
  \bibfield  {author} {\bibinfo {author} {\bibfnamefont {E.}~\bibnamefont
  {Jeckelmann}}\ and\ \bibinfo {author} {\bibfnamefont {H.}~\bibnamefont
  {Fehske}},\ }\href@noop {} {\bibfield  {journal} {\bibinfo  {journal}
  {Rivista del Nuovo Cimento}\ }\textbf {\bibinfo {volume} {30}},\ \bibinfo
  {pages} {259} (\bibinfo {year} {2007})}\BibitemShut {NoStop}%
\bibitem [{\citenamefont {Gebhard}\ \emph {et~al.}(2003)\citenamefont
  {Gebhard}, \citenamefont {Jeckelmann}, \citenamefont {Mahlert},\ and\
  \citenamefont {Noack}}]{GJMNN03}%
  \BibitemOpen
  \bibfield  {author} {\bibinfo {author} {\bibfnamefont {F.}~\bibnamefont
  {Gebhard}}, \bibinfo {author} {\bibfnamefont {E.}~\bibnamefont {Jeckelmann}},
  \bibinfo {author} {\bibfnamefont {S.}~\bibnamefont {Mahlert}, \bibfnamefont
  {S.~Nishimoto}}, \ and\ \bibinfo {author} {\bibfnamefont {R.~M.}\
  \bibnamefont {Noack}},\ }\href {\doibase 10.1140/epjb/e2004-00005-5}
  {\bibfield  {journal} {\bibinfo  {journal} {Eur.\ Phys.\ J.\ B}\ }\textbf
  {\bibinfo {volume} {36}},\ \bibinfo {pages} {491} (\bibinfo {year}
  {2003})}\BibitemShut {NoStop}%
\bibitem [{\citenamefont {Nishimoto}\ and\ \citenamefont
  {Jeckelmann}(2004)}]{Nishimoto2004}%
  \BibitemOpen
  \bibfield  {author} {\bibinfo {author} {\bibfnamefont {S.}~\bibnamefont
  {Nishimoto}}\ and\ \bibinfo {author} {\bibfnamefont {E.}~\bibnamefont
  {Jeckelmann}},\ }\href {\doibase 10.1088/0953-8984/16/4/010} {\bibfield
  {journal} {\bibinfo  {journal} {J.\ Phys.\ Condens.\ Matter}\ }\textbf
  {\bibinfo {volume} {16}},\ \bibinfo {pages} {613} (\bibinfo {year}
  {2004})}\BibitemShut {NoStop}%
 \bibitem [{\citenamefont {Paech}\ and\ \citenamefont
  {Jeckelmann}(2014)}]{PhysRevB.89.195101}%
  \BibitemOpen
  \bibfield  {author} {\bibinfo {author} {\bibfnamefont {M.}~\bibnamefont
  {Paech}}\ and\ \bibinfo {author} {\bibfnamefont {E.}~\bibnamefont
  {Jeckelmann}},\ }\href {\doibase 10.1103/PhysRevB.89.195101} {\bibfield
  {journal} {\bibinfo  {journal} {Phys. Rev. B}\ }\textbf {\bibinfo {volume}
  {89}},\ \bibinfo {pages} {195101} (\bibinfo {year} {2014})}\BibitemShut
  {NoStop}%
  \bibitem [{\citenamefont {Harris}\ and\ \citenamefont
  {Lange}(1967)}]{PhysRev.157.295}%
  \BibitemOpen
  \bibfield  {author} {\bibinfo {author} {\bibfnamefont {A.~B.}\ \bibnamefont
  {Harris}}\ and\ \bibinfo {author} {\bibfnamefont {R.~V.}\ \bibnamefont
  {Lange}},\ }\href {\doibase 10.1103/PhysRev.157.295} {\bibfield  {journal}
  {\bibinfo  {journal} {Phys. Rev.}\ }\textbf {\bibinfo {volume} {157}},\
  \bibinfo {pages} {295} (\bibinfo {year} {1967})}\BibitemShut {NoStop}%
\bibitem [{\citenamefont {van Dongen}(1994)}]{PhysRevB.49.7904}%
  \BibitemOpen
  \bibfield  {author} {\bibinfo {author} {\bibfnamefont {P.~G.~J.}\
  \bibnamefont {van Dongen}},\ }\href {\doibase 10.1103/PhysRevB.49.7904}
  {\bibfield  {journal} {\bibinfo  {journal} {Phys. Rev. B}\ }\textbf {\bibinfo
  {volume} {49}},\ \bibinfo {pages} {7904} (\bibinfo {year}
  {1994})}\BibitemShut {NoStop}%
\bibitem [{\citenamefont {Karowski}\ and\ \citenamefont {Weisz}(1978)}]{KS78}%
  \BibitemOpen
  \bibfield  {author} {\bibinfo {author} {\bibfnamefont {M.}~\bibnamefont
  {Karowski}}\ and\ \bibinfo {author} {\bibfnamefont {P.}~\bibnamefont
  {Weisz}},\ }\href@noop {} {\bibfield  {journal} {\bibinfo  {journal} {Nucl.
  Phys. B}\ }\textbf {\bibinfo {volume} {139}},\ \bibinfo {pages} {455}
  (\bibinfo {year} {1978})}\BibitemShut {NoStop}%
\bibitem [{\citenamefont {Smirnov}(1992)}]{Smirnov92}%
  \BibitemOpen
  \bibfield  {author} {\bibinfo {author} {\bibfnamefont {F.~A.}\ \bibnamefont
  {Smirnov}},\ }in\ \href@noop {} {\emph {\bibinfo {booktitle} {Form Factors in
  Completely Integrable Models of Quantum Field Theory}}}\ (\bibinfo
  {publisher} {World Scientific},\ \bibinfo {address} {Singapore},\ \bibinfo
  {year} {1992})\BibitemShut {NoStop}%
\bibitem [{\citenamefont {Controzzi}\ \emph {et~al.}(2001)\citenamefont
  {Controzzi}, \citenamefont {Essler},\ and\ \citenamefont {Tsvelik}}]{CET01}%
  \BibitemOpen
  \bibfield  {author} {\bibinfo {author} {\bibfnamefont {D.}~\bibnamefont
  {Controzzi}}, \bibinfo {author} {\bibfnamefont {F.~H.~L.}\ \bibnamefont
  {Essler}}, \ and\ \bibinfo {author} {\bibfnamefont {A.~M.}\ \bibnamefont
  {Tsvelik}},\ }\href {\doibase 10.1103/PhysRevLett.86.680} {\bibfield
  {journal} {\bibinfo  {journal} {Phys. Rev. Lett.}\ }\textbf {\bibinfo
  {volume} {86}},\ \bibinfo {pages} {680} (\bibinfo {year} {2001})}\BibitemShut
  {NoStop}%
\bibitem [{\citenamefont {Jeckelmann}\ \emph {et~al.}(2000)\citenamefont
  {Jeckelmann}, \citenamefont {Gebhard},\ and\ \citenamefont {Essler}}]{JGE00}%
  \BibitemOpen
  \bibfield  {author} {\bibinfo {author} {\bibfnamefont {E.}~\bibnamefont
  {Jeckelmann}}, \bibinfo {author} {\bibfnamefont {F.}~\bibnamefont {Gebhard}},
  \ and\ \bibinfo {author} {\bibfnamefont {F.~H.~L.}\ \bibnamefont {Essler}},\
  }\href@noop {} {\bibfield  {journal} {\bibinfo  {journal} {Phys. Rev. Lett.}\
  }\textbf {\bibinfo {volume} {85}},\ \bibinfo {pages} {3910} (\bibinfo {year}
  {2000})}\BibitemShut {NoStop}%
\bibitem [{\citenamefont {Essler}\ \emph {et~al.}(2001)\citenamefont {Essler},
  \citenamefont {Gebhard},\ and\ \citenamefont {Jeckelmann}}]{EGJ01}%
  \BibitemOpen
  \bibfield  {author} {\bibinfo {author} {\bibfnamefont {F.~H.~L.}\
  \bibnamefont {Essler}}, \bibinfo {author} {\bibfnamefont {F.}~\bibnamefont
  {Gebhard}}, \ and\ \bibinfo {author} {\bibfnamefont {E.}~\bibnamefont
  {Jeckelmann}},\ }\href {\doibase 10.1103/PhysRevB.64.125119} {\bibfield
  {journal} {\bibinfo  {journal} {Phys. Rev. B}\ }\textbf {\bibinfo {volume}
  {64}},\ \bibinfo {pages} {125119} (\bibinfo {year} {2001})}\BibitemShut
  {NoStop}%
\bibitem [{\citenamefont {Pollet}\ and\ \citenamefont
  {Prokof’ev}(2012)}]{Pollet2012}%
  \BibitemOpen
  \bibfield  {author} {\bibinfo {author} {\bibfnamefont {L.}~\bibnamefont
  {Pollet}}\ and\ \bibinfo {author} {\bibfnamefont {N.}~\bibnamefont
  {Prokof’ev}},\ }\href {\doibase 10.1103/PhysRevLett.109.010401} {\bibfield
  {journal} {\bibinfo  {journal} {Phys. Rev. Lett.}\ }\textbf {\bibinfo
  {volume} {109}},\ \bibinfo {pages} {010401} (\bibinfo {year}
  {2012})}\BibitemShut {NoStop}%
\bibitem [{\citenamefont {Teichmann}\ \emph {et~al.}(2009)\citenamefont
  {Teichmann}, \citenamefont {Hinrichs}, \citenamefont {Holthaus},\ and\
  \citenamefont {Eckardt}}]{Teichmann2009a}%
  \BibitemOpen
  \bibfield  {author} {\bibinfo {author} {\bibfnamefont {N.}~\bibnamefont
  {Teichmann}}, \bibinfo {author} {\bibfnamefont {D.}~\bibnamefont {Hinrichs}},
  \bibinfo {author} {\bibfnamefont {M.}~\bibnamefont {Holthaus}},  and\
  \bibinfo {author} {\bibfnamefont {A.}~\bibnamefont {Eckardt}},\ }\href@noop
  {} {\bibfield  {journal} {\bibinfo  {journal} {Phys. Rev. B}\ }\textbf
  {\bibinfo {volume} {79}},\ \bibinfo {pages} {224515} (\bibinfo {year}
  {2009})}\BibitemShut {NoStop}%
\bibitem [{\citenamefont {Eckardt}(2009)}]{Eckardt2009}%
  \BibitemOpen
  \bibfield  {author} {\bibinfo {author} {\bibfnamefont {A.}~\bibnamefont
  {Eckardt}},\ }\href {\doibase 10.1103/PhysRevB.79.195131} {\bibfield
  {journal} {\bibinfo  {journal} {Phys. Rev. B}\ }\textbf {\bibinfo {volume}
  {79}},\ \bibinfo {pages} {195131} (\bibinfo {year} {2009})}\BibitemShut
  {NoStop}%
\bibitem [{\citenamefont {Damski}\ and\ \citenamefont
  {Zakrzewski}(2006)}]{Damski2006}%
  \BibitemOpen
  \bibfield  {author} {\bibinfo {author} {\bibfnamefont {B.}~\bibnamefont
  {Damski}}\ and\ \bibinfo {author} {\bibfnamefont {J.}~\bibnamefont
  {Zakrzewski}},\ }\href {\doibase 10.1103/PhysRevA.74.043609} {\bibfield
  {journal} {\bibinfo  {journal} {Phys. Rev. A}\ }\textbf {\bibinfo {volume}
  {74}},\ \bibinfo {pages} {043609} (\bibinfo {year} {2006})}\BibitemShut
  {NoStop}%
\end{thebibliography}

%

\end{document}